\documentclass{article}

\usepackage{array,multirow,makecell}
\usepackage{hyperref}
\usepackage{fancybox}
\usepackage{colortbl}
\usepackage{float}

\usepackage{tikz}
\usepackage{multicol}
\usepackage{multirow}
\usepackage{booktabs}

\usepackage[utf8]{inputenc}
\usepackage{booktabs,lipsum}
\usepackage[ruled, norelsize, linesnumbered]{algorithm2e}
\usepackage{algpseudocode}

\usepackage{balance}

\usepackage{amsmath} 
\usepackage{amssymb}  

\usepackage{subfig}

\usepackage{xcolor}

\usepackage[percent]{overpic}
\usepackage{rotating}
\usepackage{cases}

\allowdisplaybreaks 

\def\aujour{\number\day \space \ifcase\month\or
janvier\or f�vrier\or mars\or avril\or mai\or
juin\or juillet\or ao�t\or septembre\or octobre\or
novembre\or d�cembre\fi \space \number\year}

\def\cH{{\cal H}}

\def\cL{{\cal L}}

\def\C{{\setbox0=\hbox{$\displaystyle{\rm C}$}
        \hbox{\hbox to0pt{\kern 0.4\wd0\vrule height 0.95\ht0\hss}\box0}}}
\def\Q{{\setbox0=\hbox{$\displaystyle{\rm Q}$}%
    \hbox{\raise 0.2\ht0\hbox to0pt{\kern 0.4\wd0\vrule height
    0.85\ht0\hss}\box0}}} 

\def\cH2{{\cal H}_2} 

\def\cL2{\mathop{\mathcal L}_{2}} 

\def\cRH2{\mathop{\cal R \cal H}_2} 
\def\cRL2{\mathop{\cal R \cal L}_{2}} 

\def\Ltwo{{\bf L}_2}
\def\Lone{{\bf L}_1}

\DeclareMathOperator*{\der}{d}

\newcommand{\abs}[1]{\left|{#1}\right|}

\makeatletter
\DeclareRobustCommand\sfrac[1]{\@ifnextchar/{\@sfrac{#1}}
                                            {\@sfrac{#1}/}}
\def\@sfrac#1/#2{\leavevmode\kern.1em\raise.5ex
         \hbox{$\m@th\fontsize\sf@size\z@
                           \selectfont#1$}\kern-.1em
         /\kern-.15em\lower.25ex
          \hbox{$\m@th\fontsize\sf@size\z@
                            \selectfont#2$}}
\makeatother

\usepackage{cite}
\usepackage{siunitx}

\usepackage{arxiv}

\usepackage[utf8]{inputenc} 
\usepackage[T1]{fontenc}    
\usepackage{hyperref}       
\usepackage{url}            
\usepackage{booktabs}       
\usepackage{amsfonts}       
\usepackage{nicefrac}       
\usepackage{microtype}      
\usepackage{lipsum}		
\usepackage{graphicx}
\usepackage[round]{natbib}
\usepackage{doi}

\newcommand{\recac}[2][]{
    \ifthenelse{\equal{#1}{def}}{x_\text{r}(t)}{x_{\text{r}#2}(t)}
    }

\title{Fish Growth Trajectory Tracking via Reinforcement Learning in Precision Aquaculture}

\author{ \href{https://orcid.org/orcid.org/0000-0003-4342-8341}{\includegraphics[scale=0.06]{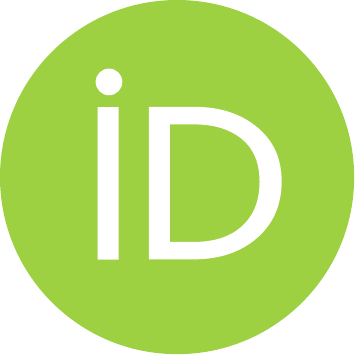}\hspace{1mm}A.~Chahid$^{1}$}, 
\href{https://orcid.org/0000-0002-2576-5515}{\includegraphics[scale=0.06]{orcid.pdf}\hspace{1mm} I. N'Doye$^{1}$}, 
\href{https://orcid.org/}{\includegraphics[scale=0.06]{orcid.pdf}\hspace{1mm}J. E. Majoris$^{2}$}, 
\href{https://orcid.org/}{\includegraphics[scale=0.06]{orcid.pdf}\hspace{1mm}M. L. Berumen$^{2}$}, 
\href{https://orcid.org/0000-0001-5944-0121}{\includegraphics[scale=0.06]{orcid.pdf}\hspace{1mm}T. -M.~Laleg-Kirati$^1$}
\thanks{This work has been supported by the King Abdullah University of Science and Technology (KAUST), Base Research Fund (BAS/1/1627-01-01) to Taous Meriem Laleg and KAUST AI Initiative.} \\
$^1$Computer, Electrical and Mathematical Sciences and Engineering Division (CEMSE)\\
King Abdullah University of Science and Technology (KAUST)\\
Thuwal 23955-6900, Saudi Arabia \\
	\texttt{abderrazak.chahid@kaust.edu.sa; ibrahima.ndoye@kaust.edu.sa; taousmeriem.laleg@kaust.edu.sa} \\
$^2$Red Sea Research Center, Biological and Environmental Science and Engineering Division\\ 
King Abdullah University of Science and Technology (KAUST)\\
Thuwal 23955-6900, Saudi Arabia \\
\texttt{john.majoris@kaust.edu.sa; michael.berumen@kaust.edu.sa}\\
}

\renewcommand{\shorttitle}{\textit{arXiv} Template}

\hypersetup{
pdftitle={A template for the arxiv style},
pdfsubject={q-bio.NC, q-bio.QM},
pdfauthor={David S.~Hippocampus, Elias D.~Striatum},
pdfkeywords={First keyword, Second keyword, More},
}

\begin{document}
\maketitle

\begin{abstract}
This paper studies the fish growth trajectory tracking via reinforcement learning under a representative bioenergetic growth model. Due to the complex aquaculture condition and uncertain environmental factors such as temperature, dissolved oxygen, un-ionized ammonia, and strong nonlinear couplings, including multi-inputs of the fish growth model, the growth trajectory tracking problem can not be efficiently solved by the model-based control approaches in precision aquaculture. To this purpose, we formulate the growth trajectory tracking problem as sampled-data optimal control using discrete state-action pairs Markov decision process. We propose two Q-learning algorithms that learn the optimal control policy from the sampled data of the fish growth trajectories at every stage of the fish life cycle from juveniles to the desired market weight in the aquaculture environment. The Q-learning scheme learns the optimal feeding control policy to fish growth rate cultured in cages and the optimal feeding rate control policy with an optimal temperature profile for the aquaculture fish growth rate in tanks. The simulation results demonstrate that both Q-learning strategies achieve high trajectory tracking performance with less amount feeding rates.
\end{abstract}


\section{Introduction}
Aquaculture is considered one of the largest and fastest-growing food production sectors worldwide. It is likely to become the primary source of seafood in the future \cite{Fao:18}. As commercial fish production increases, both its impact and reliance on ocean fisheries' protein sources are likely to expand. Hence, the efficiency of the practice, protocols, and management in the aquaculture system need to be optimized to guarantee the optimal fish growth and monitoring throughout the grow-out cycle from stocking through harvesting \cite{Seg:16}. Thus, there is a pressing need to develop new precision aquaculture techniques to improve fish farming efficiency, such as optimizing feeding and water quality control policies \cite{NGPKB:00}. This objective is achieved by optimizing factors that have a strong influence on fish growth, such as the feeding rate, temperature, and dissolved oxygen. 

Many difficulties arise for the growth trajectory tracking problem. Strong nonlinear couplings and multi-inputs of the fish growth dynamic model that linear and nonlinear model-based controllers can not handle well. Moreover, fish feeding is affected by various external factors like water temperature, dissolved oxygen, salinity, and light \cite{SHL:16}, increasing the difficulty involved with the growth trajectory tracking problem.

Modern aquaculture systems can benefit from integrating emerging technologies and theories from multiple research disciplines such as marine science and optimal control systems. In integrating control systems, classical feedback approaches are not directly convenient to most feeding regimes due to the scheduled nature of the feed ration and biological constraints of the aquaculture environment. The integration of new technology-based solutions and policies may help to promote sustainable aquaculture production. There are no examples of closed-loop precision fish farming systems, including the different components of observing the fish for decision making \cite{For:18}. Moreover, the increasing complexity of modern aquaculture systems can introduce a great deal of uncertainty in our system dynamics knowledge. These uncertainties present a significant challenge to the model-based controllers. In particular, the bioenergetic growth model is sensitive to the extrinsic environmental factors in aquaculture systems. Therefore, the aquaculture industry and researchers aspire to develop strategies that optimize biomass production by monitoring and controlling factors that influence fish growth.

To this purpose, we design a model-free reinforcement learning (RL) based controller to solve the growth trajectory tracking problem in this paper. RL is a model-free technique under the dynamic programming framework that solves the Markov decision process (MDP). It learns an optimal control policy without the complete knowledge of the dynamical model using training samples data. It has been implemented in control applications including the set-point tracking error control problems \cite{LeL:05}, the autonomous underwater vehicle (AUV) position tracking control \cite{WSYW:18,CDWMPA:18}, the robust quadrotor control \cite{WSHS:20}, the process industry under changing environment \cite{LiD:20}, etc.

This work proposes two reinforcement learning-based controllers using a Q-learning algorithm to reduce the operating feeding quantity and increase production efficiency. The proposed RL strategies learn the optimal feeding policies from the Nile tilapia's bioenergetic growth trajectories (\textit{Oreochromis niloticus}). We propose two optimal RL policies to optimize biomass production by monitoring and controlling factors that influence fish growth.


This paper is organized as follows. In Section \ref{sec-problem}, we present the general fish growth model of Nile tilapia (\textit{Oreochromis niloticus}) is presented, which describes the fish growth trajectories of the aquaculture system and discuss the optimal fish growth tracking control objective. In Section \ref{sec-MDP}, we model the growth reference trajectory tracking as MDPs to define the state and the cost. In Section \ref{sec-Q}, the RL algorithm based on the Q-learning scheme is proposed to solve the MDPs. Section \ref{num-sim} presents the obtained results and discusses the findings. Finally, concluding remarks are summarized in Section \ref{sec-dis-con}.

\section{Problem Formulation}\label{sec-problem}
In this section, we describe the representative bioenergetic growth model and discuss the optimal tracking control objective.

\subsection{Fish Growth Modeling}\label{sec-description}
A representative two-term bioenergetic fish growth model that captures the dominant growth factors, including adequate fish size, feed ration, and water temperature, is proposed in this work. The bioenergetic model is obtained from the dynamic energy budget. It presents a mechanistic basis for understanding an organism's energetics used to model the mass and energy flow through the fish from the uptake to usage for maintenance, reproduction, growth, and excretion \cite{Koo:12,LiS:08,MHGEJ:20}. The model is expressed in terms of energy fluxes between the organism and the environment. It constitutes useful tools in the early stage of an aquaculture activity to carry the capacity of a system before installing new farms \cite{VLGTH:20,FGCG:14} estimate production and feeding ration \cite{ChB:98}, or to optimize integrated multi-trophic aquaculture systems \cite{RSPFG:12}. 

According to Ursin's work \cite{Urs:67}, the fish growth model in both recirculating aquaculture systems and marine cages can be expressed as the difference between anabolism, and catabolism \cite{YLD:96,Yan:98,LIU1992,Karimanzira2016}. In this paper, a bioenergetic growth model is adopted for Nile tilapia cultured in fertilized marine ponds, incorporating available pond dynamic and fish physiology information. The model includes the effects of different parameters such as water temperature, body size, un-ionized ammonia (UIA), dissolved oxygen (DO), photoperiod, and food availability \cite{Yan:98}. Thus, the growth rate model of Nile tilapia (\textit{Oreochromis niloticus}) is described as the difference between anabolism and catabolism~\cite{Yan:98} 
\begin{equation}\label{sys1}
    \frac{\der w}{\der t}= \underbrace{\Psi\big(f, T, DO\big)v(UIA)}_{\mbox{\scriptsize anabolism}} w^m - \underbrace{k(T)}_{\mbox{\scriptsize catabolism}} w^n,
\end{equation}
where $\Psi\big(f, T, DO\big)$ (\si{g^{1\!-\!m}}\si{day^{-1}}) and $ v(UIA)$ are the coefficients of anabolism and  $k(T)$ (\si{g^{1\!-\!n}}\si{day^{-1}}) is the coefficient of fasting catabolism expressed as
\begin{equation}\label{eq1a}
\!\!\Psi\big(f, T, DO\big)= h\rho f b(1-a)\tau(T)\sigma(DO),
\end{equation}
and
\begin{equation}
k(T)=k_{\mbox{\tiny min}}\exp\Big({j(T-T_{\mbox{\tiny min}})}\Big),
\end{equation}
 where $f$ is the percent of the maximal daily ration $R_{\mbox{\tiny M}}$, defined as $R_{\mbox{\tiny M}}\!=\!3\%$ of the  body weight \cite{BHUJEL200771}. The parameters of the growth model are summarized in Table \ref{para_model} and Table \ref{para_coef}. 
\begin{table}[!t]
\caption{Nomenclature and parameters of the growth model.}
\begin{center}
\begin{tabular}{| c || c || c |}
  \hline 
\textbf{~Symbol~}  & \textbf{~~~~Description~~~~}  & \textbf{~~~~Unit~~~~}   \\ \hline  
$w$ &  Fish weight  & \si{g} \\  
~~$t$~~ & Time  & \si{day} \\
$\Psi$ &~Coefficient of net anabolism~ & ~\si{g^{1\!-\!m}}\si{day^{-1}}~ \\
$T$ &~~Temperature  &\si{^0}\si{C} \\
$DO$ &  Dissolved oxygen & \si{mg/l}   \\  
$UIA$ &~un-ionized ammonia~ & \si{mg/l} \\
$f$ &  Relative feeding rate &$0\!<\!f\!<\!1$ \\  
$k$ &~Coefficient of fasting catabolism~ &~\si{g^{1\!-\!n}}\si{day^{-1}}~ \\ \hline
~~~$m$~~~ &Exponent of body weight for net anabolism & $0.67$ \\ 
$n$ & ~~ Exponent of body weight for fasting catabolism~~  & $0.81$ \\  \hline
\end{tabular}
\end{center}
\label{para_model}
\end{table}

The effects of temperature $\tau(T)$, unionized ammonia $v(UIA)$ and dissolved oxygen $\sigma(DO)$ on food consumption are described, respectively \cite{Yan:98}.
\begin{equation*}\label{eq-temp}
\tau(T)=
\left\{\begin{array}{llll}
\displaystyle  \exp{\left\{-\kappa\Big(\dfrac{T-T_{opt}}{T_{\mbox{\tiny max}}-T_{opt}}\Big)^4\right\}} \quad \mbox{if}\quad T>T_{opt},\\
\exp{\left\{-\kappa \Big(\dfrac{T_{opt} -T}{T_{opt}-T_{\mbox{\tiny min}}}\Big)^4\right\}} \quad \mbox{if}\quad T<T_{opt},
\end{array}\right.
\end{equation*}
where $\kappa=4.6$.
\begin{equation*}\label{eq-UIA}
v(UIA)\!\!=\!\!
\left\{\begin{array}{llll}
\displaystyle 1 \qquad\qquad\qquad~\, \mbox{if} \quad UIA<UIA_{\mbox{\tiny crit}},\\
\!\!\dfrac{UIA_{\mbox{\tiny max}}-UIA}{UIA_{\mbox{\tiny max}} -UIA_{\mbox{\tiny crit}}} \, \mbox{if}\,  UIA_{\mbox{\tiny crit}}<  UIA< UIA_{\mbox{\tiny max}},\\
0 \qquad\qquad\qquad\quad \mbox{elsewhere}.
\end{array}\right.
\end{equation*}
\begin{equation*}\label{eq-DO}
\sigma(DO)=
\left\{\begin{array}{llll}
\displaystyle 1 \qquad\qquad\qquad \mbox{if} \quad DO>DO_{\mbox{\tiny crit}},\\
\!\!\dfrac{DO - DO_{\mbox{\tiny min}}}{DO_{\mbox{\tiny crit}} - DO_{\mbox{\tiny min}}}~\, \mbox{if}\quad  DO_{\mbox{\tiny min}}\!<\!DO\!<\!DO_{\mbox{\tiny crit}},\\
0 \qquad\qquad\qquad \mbox{elsewhere},
\end{array}\right.
\end{equation*}
where the different parameters of the net anabolism and fasting catabolism coefficients are summarized in Table \ref{para_coef}. The relative feeding rate $f$ is formulated as the ratio between the daily ration $d_{\mbox{\tiny r}}$ and the maximal daily ration $R_{\mbox{\tiny M}}$ as follows
\begin{equation*}
f=\dfrac{d_{\mbox{\tiny r}}}{R_{\mbox{\tiny M}}}.
\end{equation*}

\begin{table}[!t]
\caption{Parameters of the net anabolism and fasting catabolism coefficients}
\begin{center}
\begin{tabular}{| c || c || c |}
  \hline 
\textbf{~Symbol~}  & \textbf{~~~~Description~~~~}  & \textbf{~~~~Value/Unit~~~~}   \\ \hline  
$b$ &  Efficiency of food assimilation & $0.62$  \\  
$a$ & Fraction of the food assimilated & $0.53$ \\  
$h$ &Coefficient of food consumption &$0.8 $\si{g^{1\!-\!m}}\si{/day}\\   
$k_{\mbox{\tiny min}}$ &~Coefficient of fasting catabolism~ & $0.00133$\si{N}\\  
$j$ &~Coefficient of fasting catabolism~ & $0.0132$\si{N}\\  
$T_{\mbox{\tiny opt}}$ &~~Optimal average level of water temperature~~ &\si{33 ^0}\si{C}\\
$T_{\mbox{\tiny min}}$& Minimum level of temperature   &\si{24 ^0}\si{C}\\
$T_{\mbox{\tiny max}}$ &  Maximum level of temperature  &\si{40 ^0}\si{C}\\ 
$UIA_{\mbox{\tiny crit}}$ &~~~Critical limit of UIA~~~  &0.06\si{mg/l}\\
$UIA_{\mbox{\tiny max}}$ & Maximum level of UIA &1.4\si{mg/l}\\
$DO_{\mbox{\tiny crit}}$ & Critical limit of DO&0.3\si{mg/l}\\ 
$DO_{\mbox{\tiny min}}$ & Minimum level of DO &1\si{mg/l}\\ 
$d_{\mbox{\tiny r}}$ &  daily ration  &\si{g/day}\\
$R_{\mbox{\tiny M}}$ & Maximal daily ration \cite{BHUJEL200771} & $3 \% ~BWD$\\
$BWD$ & Average body-weight per day &\si{g/day}\\ \hline  
$\tau$ &~Temperature factor~ &$0\!<\!\tau\!<\!1$ \\ 
$\sigma$ &  Dissolved oxygen factor  & $0\!<\!\sigma\!<\!1$ \\  
$v$ &  un-ionized ammonia factor & $0\!<\!v\!<\!1$ \\  
$\rho$ & Photoperiod factor & $0\!<\!\rho\!<\!2$\\  \hline  
\end{tabular}
\end{center}
\label{para_coef}
\end{table}

\subsection{Objective of This Work}\label{subsec-objective}
\begin{figure}[!t]
	\centering
\includegraphics[width=0.58\linewidth]{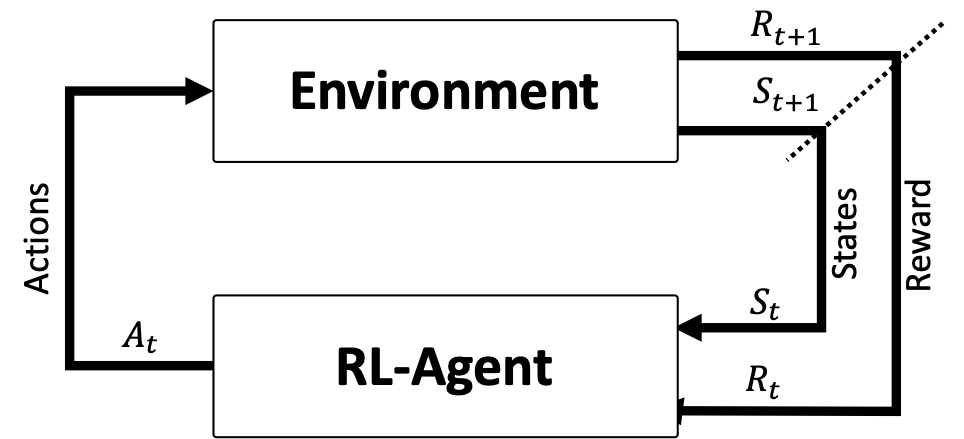}
	\caption{Reinforcement learning scheme.}
	\label{fig-RL0}
\end{figure}

The main idea behind the reinforcement learning (RL) algorithm is to understand how to choose an action at the current state to increase the final reward or the ultimate target. 
RL-based control can be a good alternative to the classical model-based control techniques as it does not need a mathematical model to find the best control strategy. It derives the control insights from the interaction with the real environment directly, as illustrated in Fig. \ref{fig-RL0}.

\begin{figure}[!t]
\centering
      \includegraphics[width=0.58\linewidth]{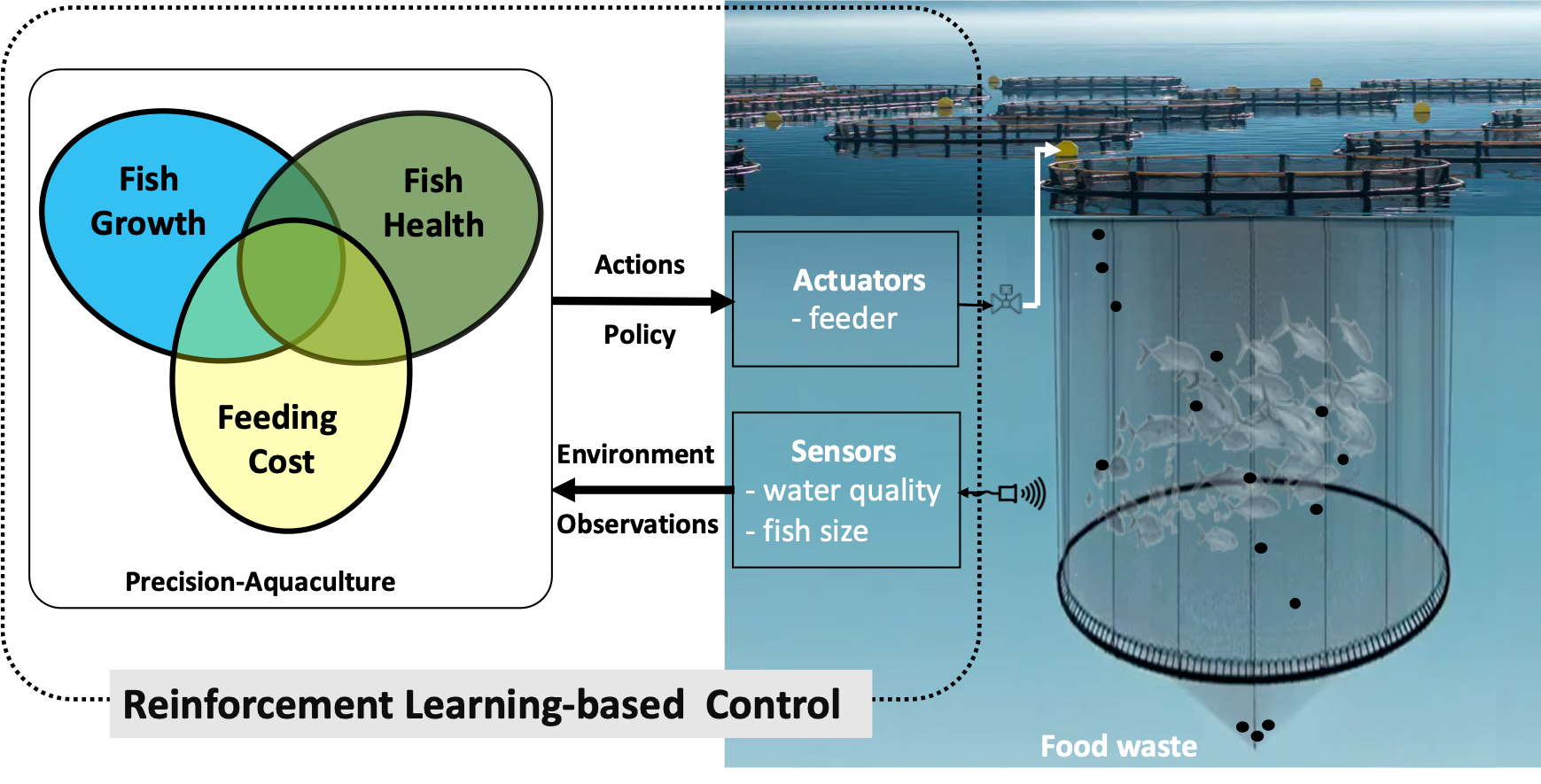}
      \caption{RL-based feeding control to fish growth rate in precision aquaculture.}\label{fig-precision_aquaculture}   
 \end{figure}

The optimal control problem studied in this paper consists of minimizing the growth tracking error deviation while penalizing the feeding ration with an optimal temperature profile for aquaculture fish tanks and the feeding rate to fish cultured in cages, respectively. We propose a model-free reinforcement learning control algorithm that learns the optimal control policies in both floating cages and tanks aquaculture environments from the growth trajectories sampled data, as illustrated in Fig. \ref{fig-precision_aquaculture}. 

The benefits of the RL scheme to address the growth trajectory tracking problem include
\begin{itemize}
  \item The bioenergetic fish growth dynamical model based on an energy budget equation is hard to derive in practice. In contrast, the RL algorithm is based on samples that do not require the growth model's knowledge.
  \item The resulting fish growth dynamical model is strongly nonlinear coupled, including multi-inputs that model-based controllers can not handle correctly.
  \item The complex aquaculture condition and the external factors, including human management operations and environment, photoperiod, un-ionized ammonia, salinity, are not restrained by the RL algorithms.
\end{itemize}

\section{Markov Decision Processes (MDP) for Fish Growth Trajectory Tracking}\label{sec-MDP}
In this section, we model the growth trajectory tracking problem as MDP to accommodate the environmental changes of the related variables and the complex bioenergetic dynamical model of fish growth in the aquaculture environment.
%

\begin{figure}[!t]
	\centering
\includegraphics[width=0.72\linewidth]{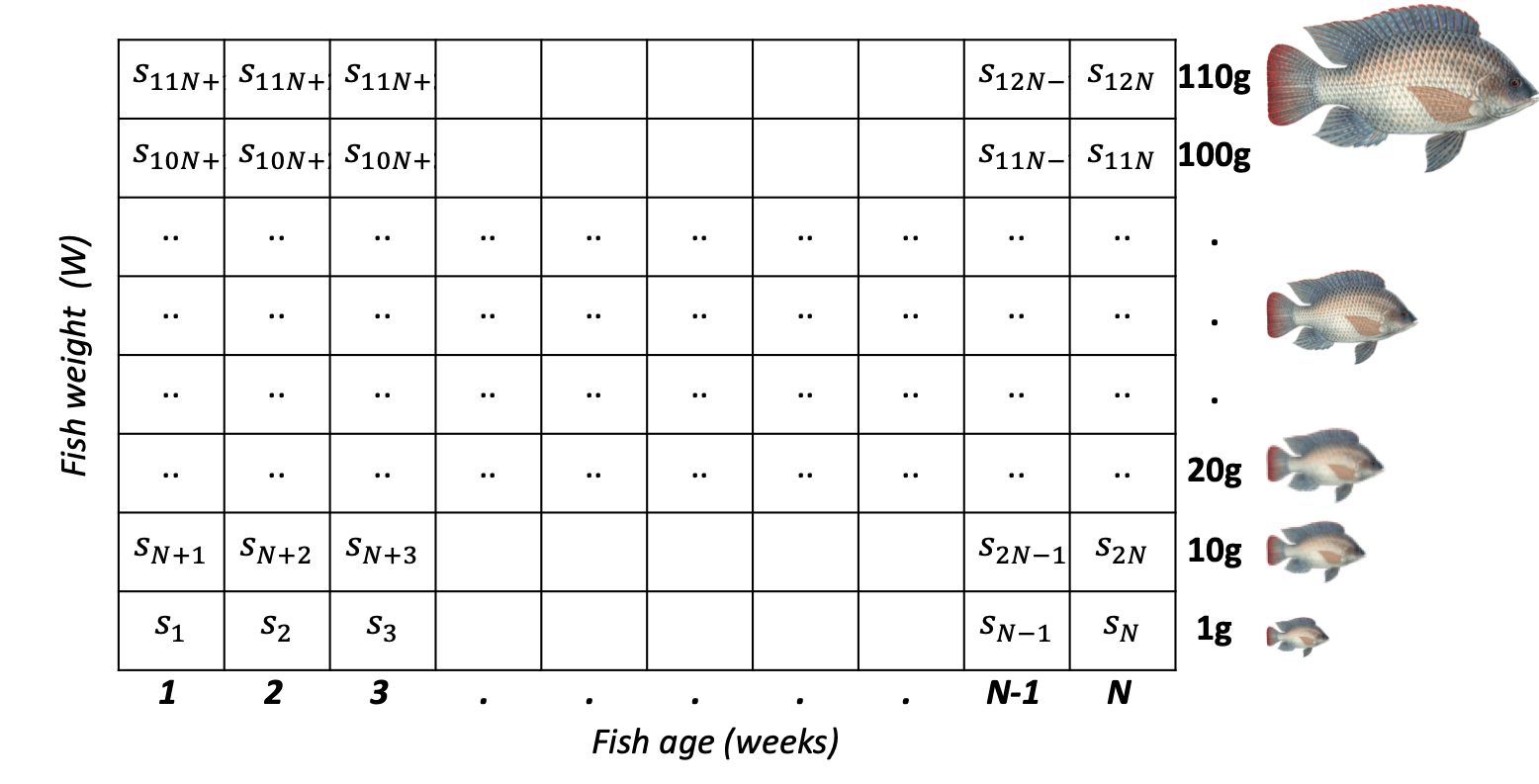}
	\caption{Example of state discretization scheme of the pair (weight, age)=$(w,t)$ with a wight resolution $\Delta w= 10$ {\si grams} and time resolution $\Delta t=7$ {\si days}, respectively.}
	\label{fig-discritization}
\end{figure}

Markov Decision Process (MDP) is used to model the aquaculture environment in the RL-like environment. This Markov property that is derived from the RL-like environment considers that the current growth state of an agent possesses all the information \cite{sutton1998r,BeT:96,watkins1992q,Bel:03,Ber:05,Pow:07,Sug:15}.
The RL-environment that is described by finite states MDP of the fish growth trajectories is defined as follows:

\noindent{\textbf{State}:} $\mathbf{S}$ is a set of finite states that describes the status of the aquaculture process. It contains parameters describing the characteristics of the aquaculture system, such as fish growth or environmental conditions. In other words, the states define the possible responses of the environment to the possible input actions such as fish weight, age, etc. To build a finite state set defined as $\mathbf{S}=\{s_0, s_1,\cdots,s_{M-1}\}$ of MDP as a lookup table, we use a discretization scheme to convert the continuous parameters of the environment to a finite set as illustrated in Fig. \ref{fig-discritization}. The states are defined by the pair (weight, age). We discretize the continuous fish growth trajectories to the fish age and adopt an approximation solution method for the feeding and temperature profiles from the growth rate model of Nile tilapia (\textit{Oreochromis niloticus}) described in \eqref{sys1}.

\begin{figure*}[!t]
    \centering
    \begin{minipage}{0.49\textwidth}
        \centering
        \includegraphics[width=0.88\textwidth]{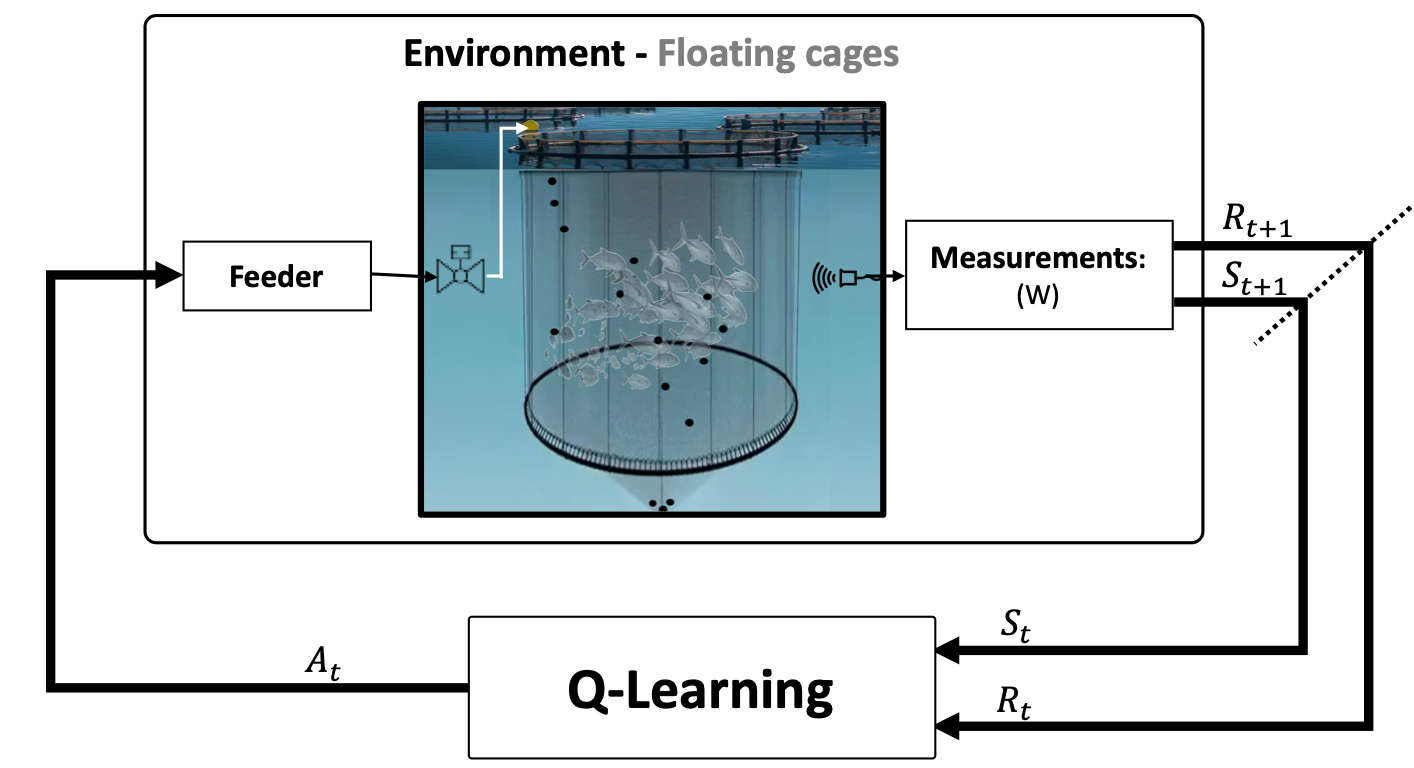}
    \end{minipage}\hfill
    \begin{minipage}{0.49\textwidth}
        \centering
        \includegraphics[width=0.88\textwidth]{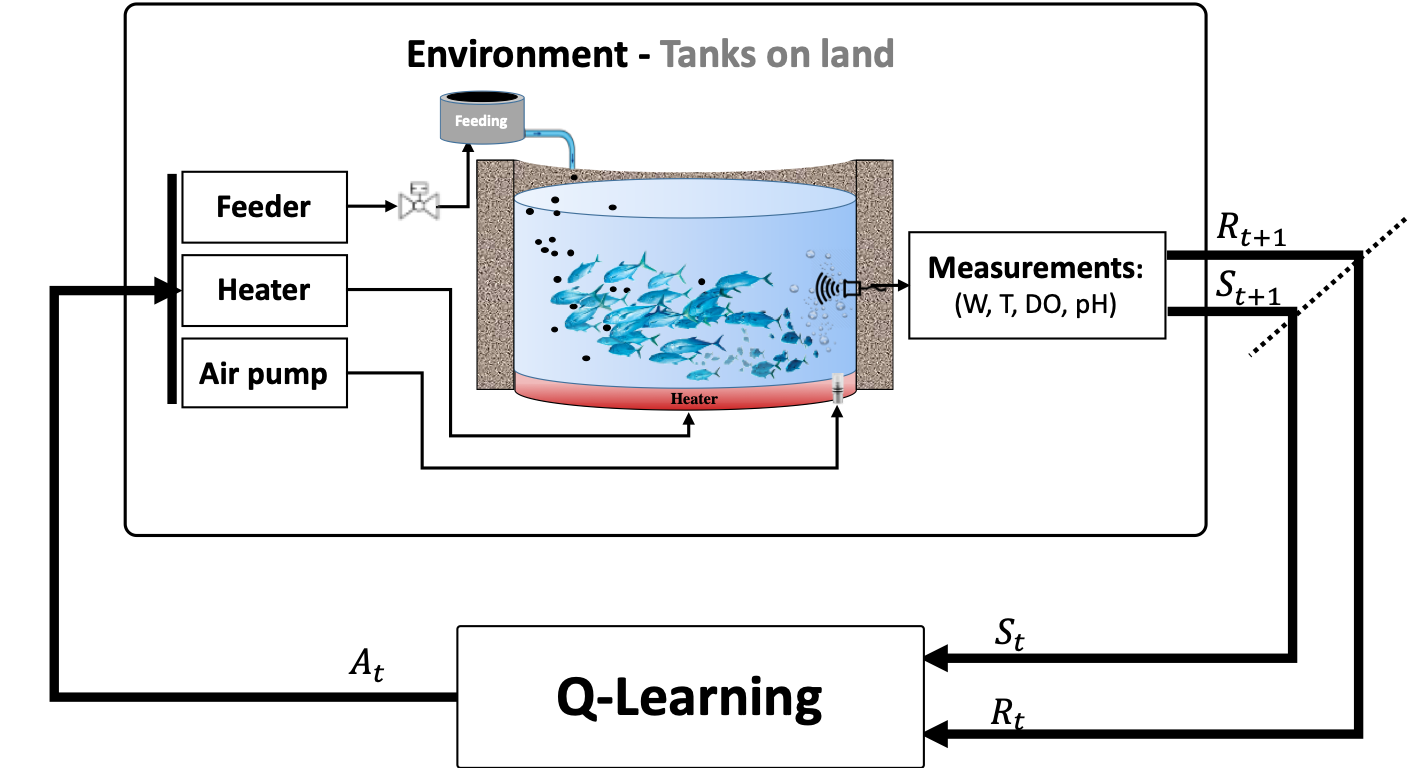}
    \end{minipage}
   \caption{Reinforcement learning framework: \textbf{(a)} fish growth in floating cages; \textbf{(b)} fish growth in tanks on land.} \label{fig:Frame_Work_RL2}
     \vspace{0.5cm}
\end{figure*}

\noindent{\textbf{Action:}} $a \in \mathbf{A_s}$ is a set of possible control actions from state $s$ representing a specific fish weight. The finite space of action is defined as $\mathbf{A_s}\subset \mathbb{R^+}$. As fish farmers can use two main type of systems to grow their fish, whether floating cages in the ocean or tanks on land (see Figs. \ref{fig:Frame_Work_RL2}\textbf{(a)} and \ref{fig:Frame_Work_RL2}\textbf{(b)}, respectively). We define two set of action $A$ as illustrated in Table \ref{tbl:RL1_RL2_actions_states}. For any trajectory generated in the state-action space, the agent follows a deterministic policy with probability equal to one.

\begin{table}[!t]
\vspace{0.5cm}
\centering
{
\caption{States/action formulation in precision aquaculture}
\label{tbl:RL1_RL2_actions_states} 
\begin{tabular}{|l|c|c|}
\hline
\multicolumn{1}{|c|}{\textbf{System}} &~~\textbf{States} $(s)$~~   & ~~\textbf{Action}  $(a)$~~  \\ \hline
~~ Floating cages~~        & ~~$(w,t)$~~          & ~~$\big(f\big)$~~               \\ \hline
~~Tanks on land~~           & ~~$(w,t)$~~           & ~~$\big(f,T\big)$~~            \\ \hline
\end{tabular}
}
\end{table}

\noindent{\textbf{Reward:}} $r(s,s')$ is reward received after transitioning from state $s$ to state $s'$, after applying the action $a$. In this work, the optimal RL policy based on MDP tracks a desired fish growth trajectory while penalizing the feed ration $\big(f\big)$ for aquaculture cages and aquaculture fish tanks in which the water temperature is controlled, respectively.

\subsection{Reference Growth Trajectory Tracking with Minimum Feeding Rate Consumption}
The first reward formulation tracks the desired fish growth trajectory while minimizing the feed ration for aquaculture fish floating cages. We formulate the reward function $r\Big(s,a\Big)$ as follows
\begin{equation}\label{reward2a}
r_t\Big(s_t,a_t\Big)= -\left[\Big(\dfrac{ w(s_t)- w^d(t)}{w^d(t)} \Big)^2 + \lambda \Big(f_{a_{t}}\Big)^2 \right],
\end{equation}
where $w(s_t)$ is the fish weight at the state $s_t$ and  $w^d(t)$ is the desired reference live-weight growth trajectory. The potential growth rate profile $w^d(t)$ is based on experimental data analysis and describes the rate achieved by a specific strain that satisfies all the nutritional requirements. $f_{a_{t}}$ is the feeding rate and $\lambda$ is a positive regularization term to assess the feeding input preference. It is tuned empirically such that a good compromise between the tracking growth error performance and the feeding consumption over the admissible space of policies.

\subsection{Reference Growth Trajectory Tracking with Minimum Feeding and Energy Consumption}
Here, the RL framework is proposed for fish growth trajectories based on water temperature control in aquaculture fish tanks. Hence,
the second reward formulation minimizes the growth tracking error deviation while penalizing the feed ration with an optimal water temperature profile.  Similarly, the reward function $r\big(s,a\big)$ is formulated as \eqref{reward2a}.

\section{Q-Learning Algorithm with Temporal Difference Update}\label{sec-Q}
To solve the MDP growth trajectory tracking problem, we propose a Q-learning algorithm based on the temporal difference method that learns from raw experience without the complete knowledge of the fish growth dynamical model in the aquaculture environment. Q-learning is one of RL's most important advances that searches for the optimal control policy using $Q(s,a)$ function \cite{watkins1992q}. This function defined the action overall value/weight/significance of each possible action $a$ at a specific states $s$. In this work, we substitute the action $a$ in the growth trajectory tracking problem with the feeding rate $(f)$ to fish cultured in floating cages and tanks land with an optimal temperature profile, respectively. At each time $t$, the weights learn from the aquaculture environment's response, then the temporal difference (TD) method that uses sampling experiences updates the action-value function.

\subsection{Q-Learning for Fish Growth Rate in Cages} 
The first action-value function based on the temporal difference (TD) principle at each time $t$ for fish growth rate in cages is defined as follows
\begin{align}\label{QRL}
Q\Big(s_{t},a_{t}\Big)\!\leftarrow\!Q\Big(s_{t},a_{t}\Big)+\alpha\left[ r_t\Big(s_t,a_t\Big)\!+\!\gamma \max _{a} Q\Big(w_{t+1},a_{t+1}\Big)\!-\!Q\Big(w_{t},a_t\Big)\right],
\end{align}
where $Q\Big(s_{t},a_{t}\Big)$ is the value function of the state-action pair $\Big(s_t,a_t\Big)$ at each time $t$ and  $r_t\Big(s_t,a_t\Big)$ is the corresponding reward for fish growth rate in cages. $\alpha$ is the learning rate and $\gamma$ is the discount factor.  Fig. \ref{fig:Frame_Work_RL2}\textbf{(a)} illustrates the RL based optimal feeding control policy for fish growth rate in cages.

\subsection{Q-Learning for Fish Growth Rate in Tanks} 
Similarly, the second action-value function based on TD principle at each step $t$ for fish growth rate in a controlled water temperature tank is defined as \eqref{QRL}.
Fig. \ref{fig:Frame_Work_RL2}\textbf{(b)}  illustrates the RL framework for optimal fish feeding control policy with an optimal temperature control profile.


\subsection{Off-Policy Learning Algorithm}
The proposed Q-learning algorithm is implemented using the exploration/exploitation scheme to achieve a globally optimal policy $\pi^*$ to consider the hidden variabilities and avoid locally optimal policies. As we are simulating the environment, excessive exploration is allowed to learn the generated data thoroughly. A sub-optimal policy will then be used to initialize the on-policy learning for real experiments while limiting the exploration phase to avoid harming fish. To smoothly alternate from the exploration to the exploitation phases, we introduce the greedy parameter $\epsilon$, which decays exponentially with the increase of the training episodes $i$ using the following thermal annealing process formula
\begin{equation}\label{greddy_annealing}
\epsilon = 1-\epsilon_0 \exp\Big(\dfrac{i}{t_\epsilon} \Big),
\end{equation}
where $i$ is the current training episode, and $t_\epsilon$ defines the exploration phase duration and the beginning of the exploitation phase. It is worth to mention that the higher the $t_\epsilon$  is, the slower the convergence of the RL-training becomes.  Algorithm \ref{algo1} provides the pseudocode to implement the policy iteration algorithm based on the Q-learning approach. The proposed algorithm is off-policy as the agent acts in the aquaculture environment according to $\epsilon$-greedy and learns its policy.
\begin{figure}[!t]
		\begin{algorithm}[H]
			\SetAlgoLined
			\SetKwInOut{Input}{input}\SetKwInOut{Output}{output}
			\Input{$w_0$: initial fish weight\\ $w^d$: reference trajectory fish weight\\ $\mathbf{S}$: set of possible growth states\\ $\mathbf{A_s}$: set of possible actions\\ $\epsilon_0$: initial greedy policy (exploration factor)\\ $t_\epsilon$: exploration phase duration}
			\Output{optimal control policy: $\pi^*$}
			$\diamond$  Initialize policy\;
			$Q(s,a)\leftarrow 0$,   \quad $\pi=\pi_Q$\quad ($\pi_Q$ is derived from $Q$)\;
			\While{stop = 0}{$\diamond$ Policy improvement\;
			$s\leftarrow 0$,  \quad $\pi_{old}=\pi$\;
			\While{ $s \neq$ `Terminal'}{
			         $\checkmark$ compute $\epsilon$ from \eqref{greddy_annealing}\;
			         $\checkmark$ choose an action $a$ from state $s$ using $\pi$ ($\epsilon$- greedy)\;
			         $\checkmark$ observe the new state $s'$ and reward $r$\;
			         $\checkmark$ $s\leftarrow s'$, \quad  $episode \leftarrow ~episode+1$\;
			         $\diamond$ Update Q-table and policy $\pi_Q$\;
				$\delta=r\big(s,a\big)+\gamma \max_{a} Q\big(s',a'\big)\!-\!Q\big(s,a\big)$\; 
				$Q\big(s,a\big) \leftarrow Q\big(s,a\big)+\alpha \delta$\;
				$\pi =\pi_Q$\;
				$\diamond$ check the stopping criteria after each episode\;
				\If{ $\pi_{old}=\pi$}{ $stop \leftarrow  1$}
			}
			$\pi^* \leftarrow  \pi$
			}
			\caption{Policy iteration based Q-learning}\label{algo1}
		\end{algorithm}
\end{figure}

\section{Numerical Simulations}\label{num-sim}
In this section, we design the optimal RL controller that minimizes the growth tracking error deviation while penalizing the feeding ration to fish cultured in cages, and the feeding rate with an optimal temperature profile for aquaculture fish tanks, respectively. The parameters of the Nile tilapia growth model are set based on the values provided in \cite{Yan:98}. Besides, the growth reference tracking profile $w^d$ is based on experimental data analysis and describes the rate achieved by a specific strain that satisfies all the nutritional requirements \cite{Dampin2012}. The bioenergetic fish growth model \eqref{sys1} with the parameters' values listed in Table \ref{para_coef} is only used to generate episodes for simulation purposes. The dissolved oxygen $DO$ level is set to its critical value $DO\!=\!0.3${\si mg/l}. The parameters listed in Tables \ref{table1} and \ref{table2} are used to implement Algorithm \ref{algo1}.

\newcolumntype{g}{>{\columncolor{gray}}c}
\vspace{0.15cm}
\begin{table}[!t]\caption{Parameters used in Algorithm \ref{algo1}}\label{table1}
\centering
\begin{tabular}{|c|g|g|}
\hline
\rowcolor{white}
~Parameter~ &~Symbol~ &~Value~ \\ \hline 
~Initial greedy policy~ & $\epsilon_0$ & $0.9$    \\ \rowcolor{white}
Discount factor (Cages and Tanks) & $\gamma$ & $0.8$    \\
Learning rate (Cages and Tanks) & $\alpha$ & $0.1$    \\\rowcolor{white}
~Regularization term (Cages and Tanks)~ & ~~$\lambda$~~ & $0.5$   \\
\hline
\end{tabular}
\vspace{0.25cm}
\end{table}

\newcolumntype{g}{>{\columncolor{gray}}c}
\vspace{0.15cm}
\begin{table}[!t]\caption{Training parameters for the trajectory tracking performance of the RL}\label{table2}
\centering
\begin{tabular}{|c|g|g|}
\hline
\rowcolor{white}
~Parameter~ &~Symbol~ &~Value~ \\ \hline 
~Episodes (Cages)~ &   & $15000$    \\ \rowcolor{white}
~Episodes (Tanks)~ &   & $30000$    \\
Growth state space & $\mathbf{S}$ & $[6, \, 400]$    \\ \rowcolor{white}
Feeding action space & $\mathbf{A_s}^{(.)}$ & ~$ [0.01,\, 1] $~  \\ 
~~Temperature action space~~ & $\mathbf{A_s}^{(.)}$ & ~$[29.6, \, 30.7]$~   \\ \rowcolor{white}
\hline
\end{tabular}
\vspace{0.15cm}
\end{table}

\subsection{Trajectory Tracking Performance of the RL Controller}\label{sub-sec3a}
We compare the tracking performance of the floating cages and tanks on land aquacultures systems illustrated in Fig.\ref{fig:Frame_Work_RL2}. 

\subsubsection{Floating Cages Environment}
\begin{figure*}[!t]
    \centering
    \begin{minipage}{0.49\textwidth}
        \centering
       \includegraphics[width=0.89\textwidth]{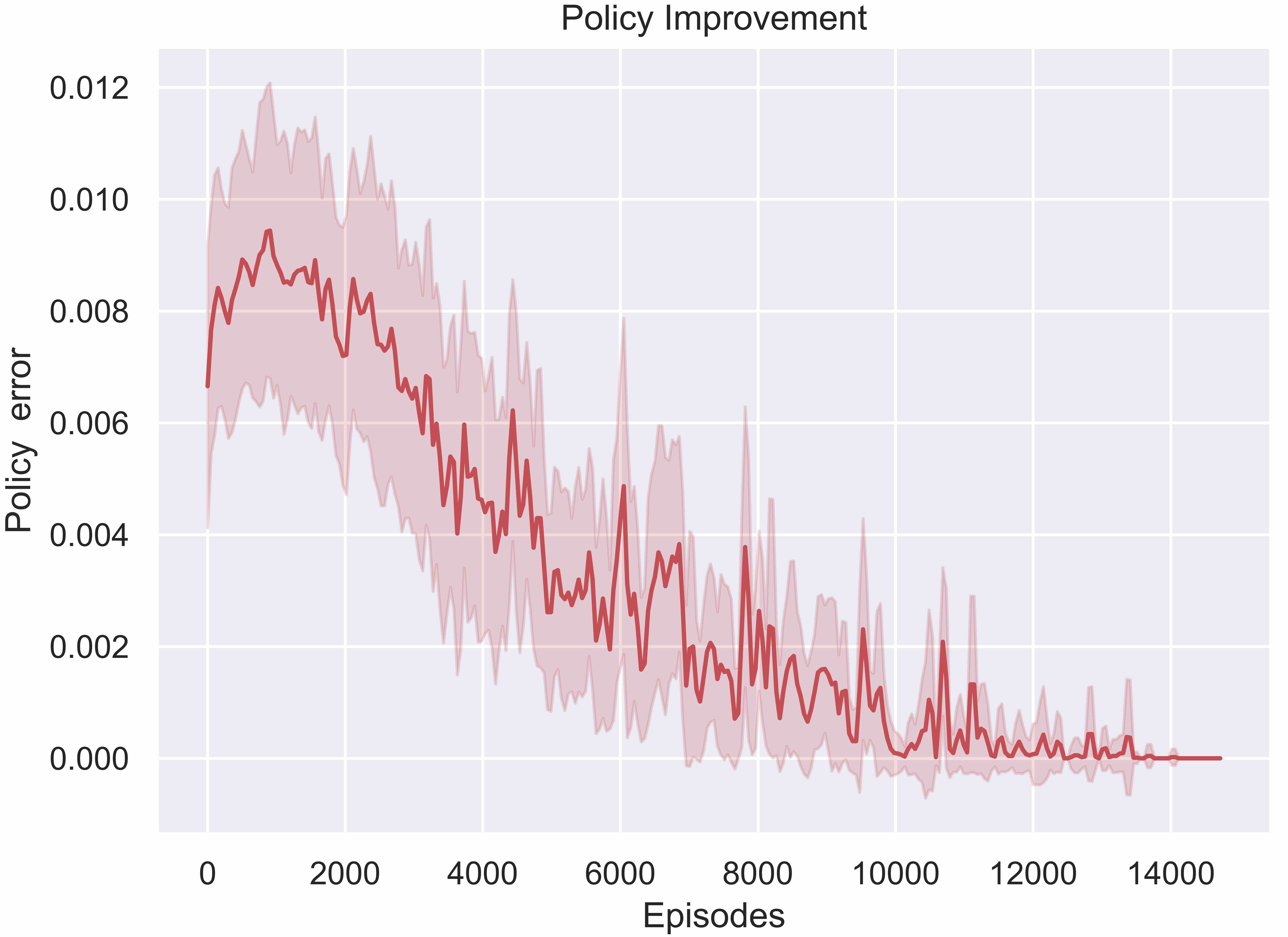}
    \end{minipage}\hfill
    \begin{minipage}{0.47\textwidth}
        \centering
       \includegraphics[width=0.98\textwidth]{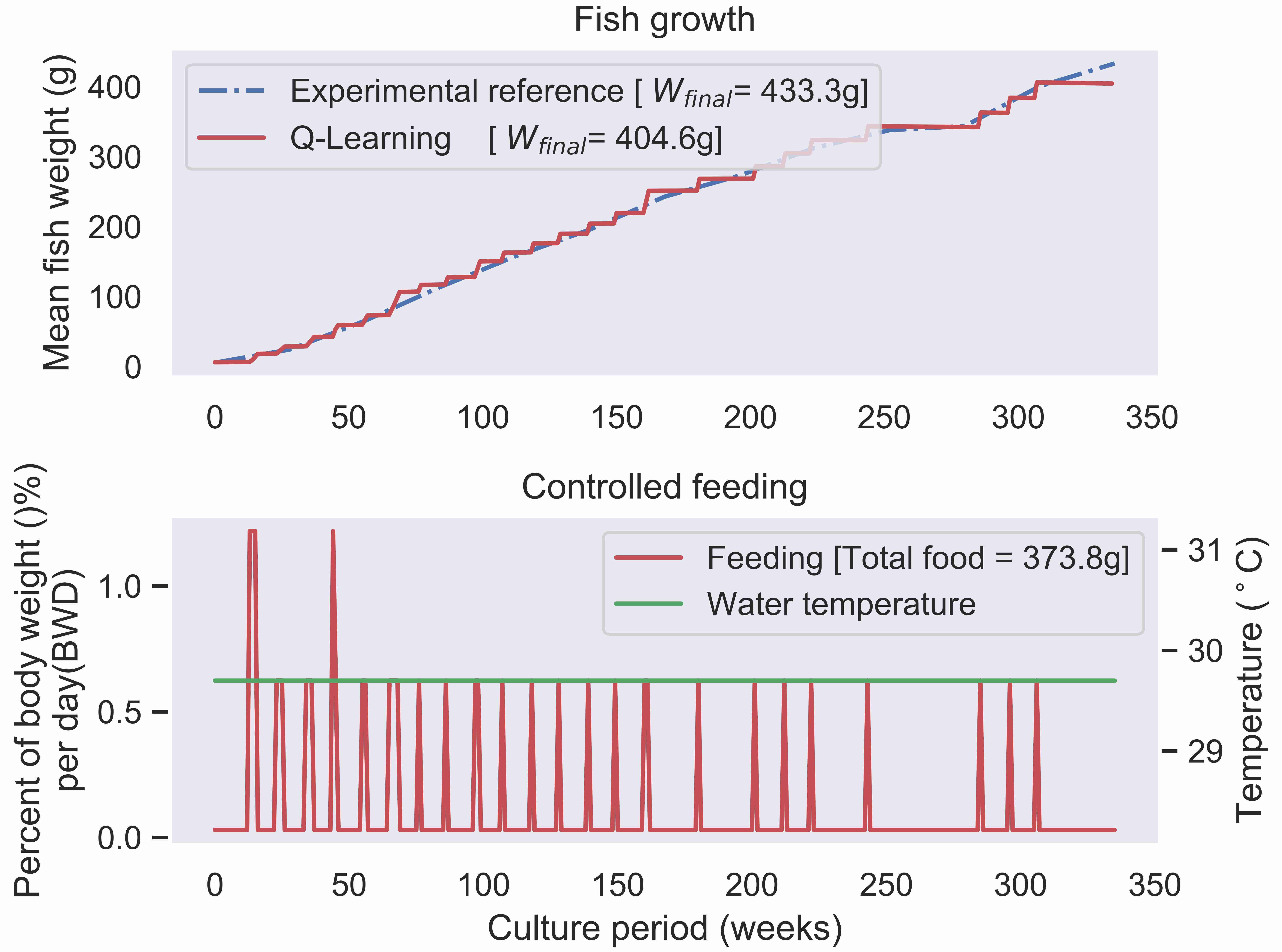}
        \end{minipage}
   \caption{Floating cages environment: \textbf{(a)} policy improvement during the training episodes; \textbf{(b)} Q-learning based feeding control strategy in which the temperature is maintained at $T\!=\!29.7^0$C.} \label{fig:policy_improvment}
\end{figure*}

Figs. \ref{fig:policy_improvment}\textbf{a)} and \ref{fig:policy_improvment}\textbf{b)} illustrate the policy improvement during the training episodes and the Q-learning based feeding control strategy in which the temperature is maintained at $T\!=\!29.7^0$C for floating cages aquaculture, respectively. From Fig. \ref{fig:policy_improvment}\textbf{a)}, we observe that the proposed RL policy interacts with the aquaculture process. Subsequently, the Q-learning policy training/iteration needs more than $10000$ episodes to converge to an optimal feeding policy. Fig. \ref{fig:policy_improvment}\textbf{b)} shows that the optimal RL controller tracks the prescribed experimental fish growth profile with minimum food. However, the RL controller operates out of the feeding action space to track the desired growth trajectory. Furthermore, it is worth mentioning that the exploration duration $t_{\epsilon}$ should be long enough to ensure minimal learning and converging to an acceptable policy as shown in Table \ref{tab-eps}. 

\subsubsection{Tanks on Land Environment}
\begin{figure*}
    \centering
    \begin{minipage}{0.44\textwidth}
        \centering
       \includegraphics[width=1\textwidth]{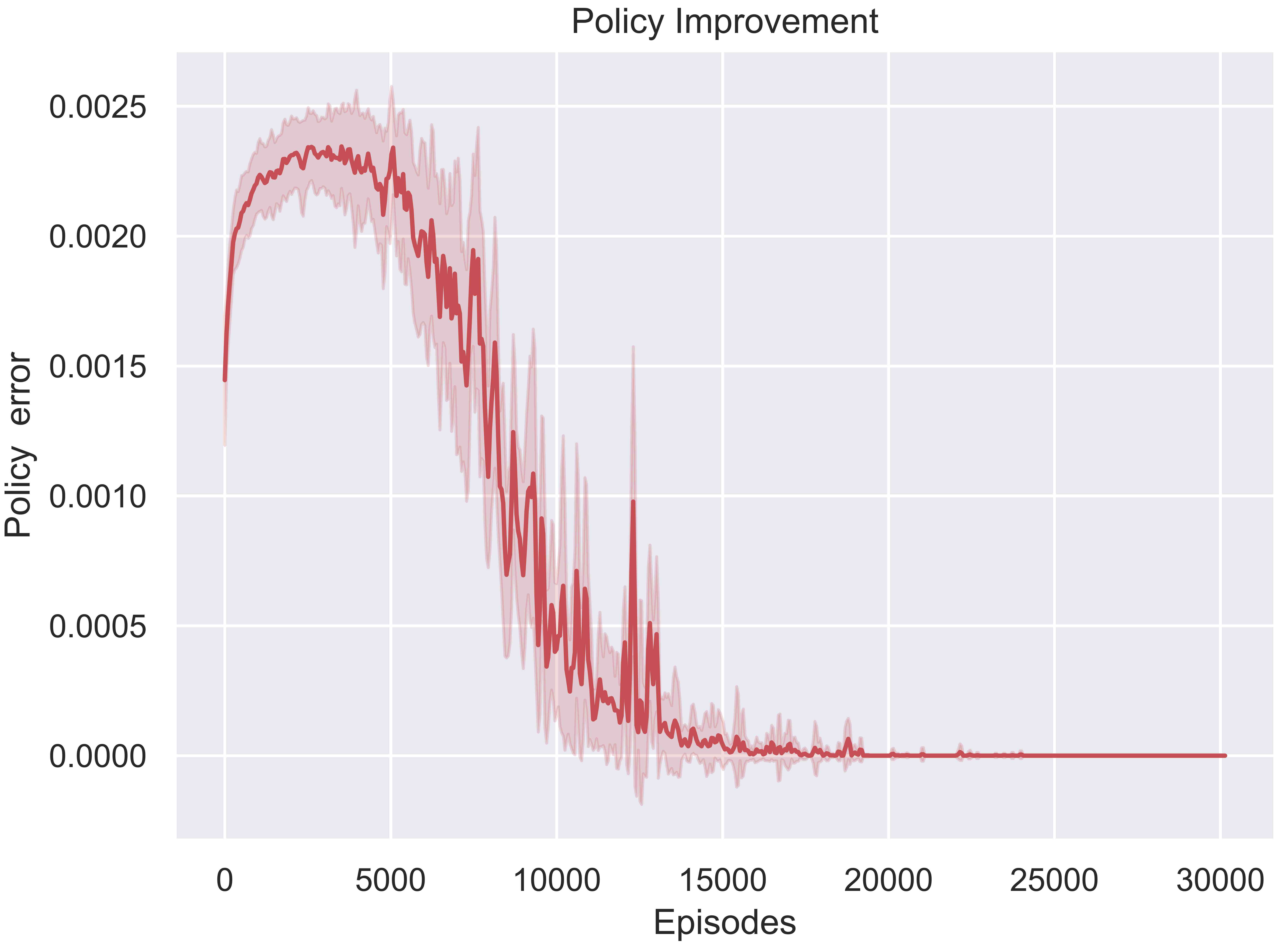}
    \end{minipage}\hfill
    \begin{minipage}{0.47\textwidth}
        \centering
      \includegraphics[width=1\textwidth]{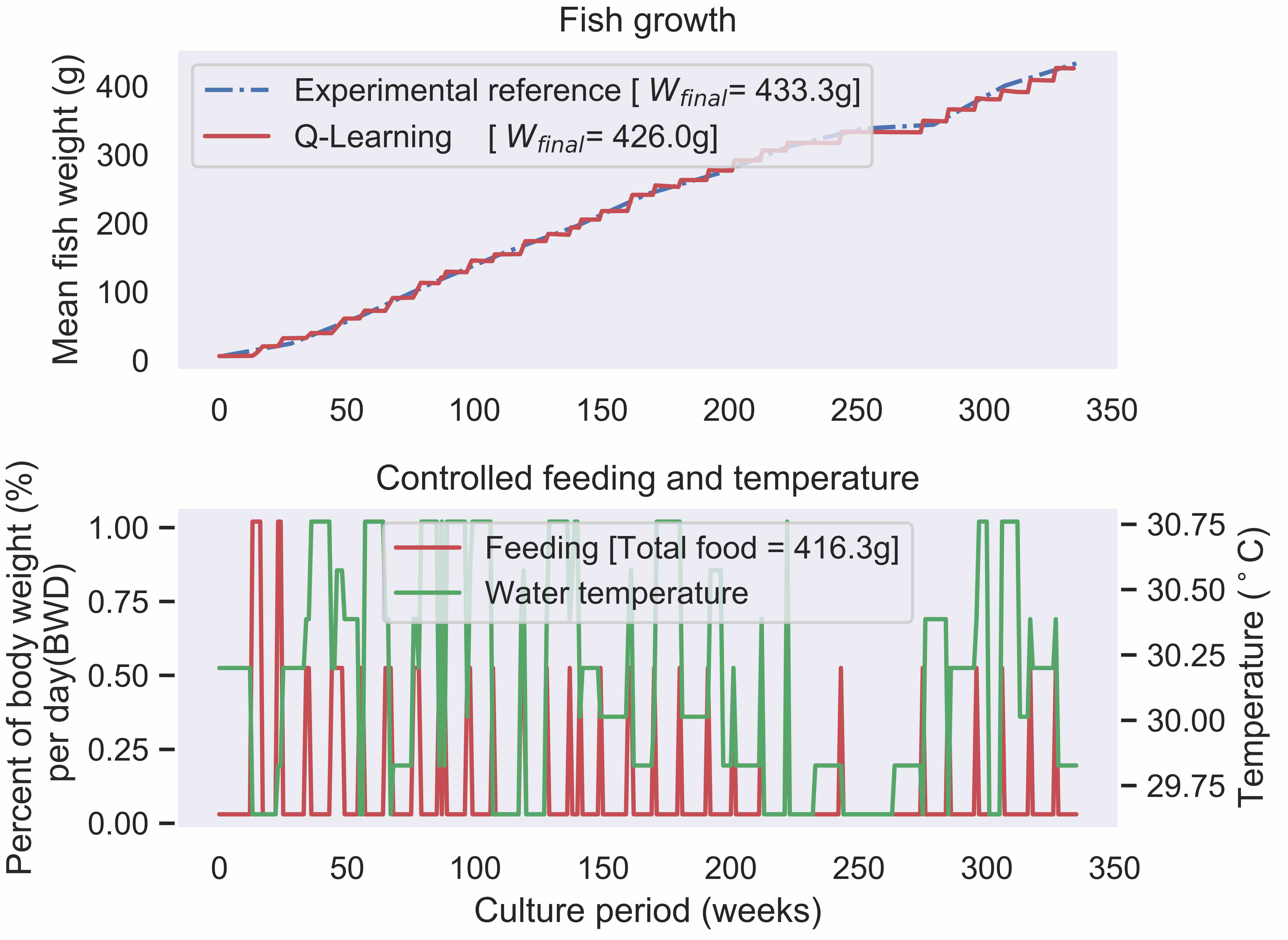}
    \end{minipage}  
   \caption{Tanks on land environment;  \textbf{(a)} policy improvement during the training episodes; \textbf{(b)} Q-learning based feeding control policy with controlled water temperature for tanks on land.} \label{fig:growth_tracking} 
    \vspace{0.75cm}
\end{figure*}

Figs. \ref{fig:growth_tracking}\textbf{a)} and \ref{fig:growth_tracking}\textbf{b)}  show that desired experimental fish growth tracking by controlling both the feeding and temperature.  It is clear that \ref{fig:growth_tracking}\textbf{b)}, the RL control policy could achieve better tracking performance and operates inside the feeding and temperature actions space. However, it consumed a slightly higher feeding rate; the extra feeding is used to compensate for the optimal temperature control error. 

The proposed RL results based on tanks on Land and floating cages aquaculture environment are encouraging and promising. They show that the Q-learning algorithm can help understand complex systems without an explicit model or define optimal operating parameters such as the optimal temperature. 
 
\subsection{Sensitivity Analysis of the Learning Parameters  (Tanks on Land Environment Case Study)}\label{sensitivity}
To compare the performance of the proposed RL control strategies, we define the following evaluation metrics as follows 
\begin{itemize}
\item \textbf{Feed conversion ratio (FCR)}
\begin{equation}
   \mbox{FCR}=\dfrac{\mbox{total feed quantity (kg)}}{\mbox{final weight (kg)} - \mbox{initial weight (kg) }},
\end{equation}
The FCR describes the quantity of feed used to the fish organisms under satisfactory conditions for its development. 
\item \textbf{Mean absolute percentage error (MAPE)}
\begin{equation}
    \mbox{MAPE}=\frac{1}{N}\sum_{i=1}^{N}{ \frac {\Big|w(i) -w^d(i)\Big|}{w(i)}}
\end{equation}
\item \textbf{ Mean absolute  error (MAE)}
\begin{equation}
    \mbox{MAE}=\frac{1}{N} \sum_{i=1}^{N}{\Big|w(i) -w^d(i)\Big|}
\end{equation}
\item \textbf{Root mean squared error (RMSE)}
\begin{equation}
\mbox{RMSE} = \sqrt{\frac{1}{N}\sum_{i=1}^{N}{\Big(w(i) -w^d(i)\Big)^2}},
\end{equation}
\end{itemize}
where $w$ and $w^d$ are the achieved and desired fish growth, respectively. $N$ is the number of culture days.

\begin{table*}[!t]

\centering
\caption{Effect of the reward and discretization scheme on the tracking performance.}
\label{descritsation_resolution}
\begin{tabular}{|c|c|c|c|c|c|c|c|c|c|}
\hline
\textbf{\begin{tabular}[c]{@{}c@{}} $\textcolor{white}{\Delta w}$~~ \\~~(grams)~~~\end{tabular}} & \textbf{\begin{tabular}[c]{@{}c@{}}$\Delta t$\\ ~~(days)~~~\end{tabular}} & \textbf{~~Reward~~} & \textbf{\begin{tabular}[c]{@{}c@{}}~Training~ \\ ~~episodes~~~\end{tabular}} & \textbf{~~MAPE (\%)~~} & \textbf{~~MAE}~~  & \textbf{~~RMSE~~} & \textbf{\begin{tabular}[c]{@{}c@{}}~~~Total food~~~~\\ ~(grams)~\end{tabular}} & \textbf{\begin{tabular}[c]{@{}c@{}}~Final fish~\\~~~weight (grams)~~~~\end{tabular}} & \textbf{~~~FCR~~~}  \\ \hline
\textbf{10}                                                                              & \textbf{7}                                                                         & $\Ltwo$           & \textbf{6382}                                                                  & \textbf{4.86}              & \textbf{4.54}          & \textbf{5.73} & \textbf{436.17}                                                                & \textbf{417.19 }                                                                           & \textbf{1.05} \\ \hline
10                                                                              & 7                                                                         & $\Ltwo\&\Lone$        & 7298                                                                  & 11.67             & 22.5          & 28.62         & 370.51                                                                & 359.38                                                                            & 1.03          \\ \hline
10                                                                     & 7                                                                & $\Lone$ & 7740                                                         & 5.01     & 4.88 & 6.13 & 445.89                                                       & 425.36                                                                   & \textbf{1.05} \\ \hline
10                                                                              & 10                                                                        & $\Ltwo$           & 4922                                                                  & 6.09              & 4.94          & 6.23          & 443.58                                                                & 425.09                                                                            & 1.04          \\ \hline
10                                                                              & 10                                                                        & $\Ltwo\&\Lone$        & 6563                                                                  & 12.87             & 23.43         & 32.46         & 362.91                                                                & 350.68                                                                            & 1.03          \\ \hline
10                                                                              & 10                                                                        & $\Lone$           & 5779                                                                  & 6.83              & 5.2           & 6.42          & 434.21                                                                & 415.4                                                                             & \textbf{1.05} \\ \hline
15                                                                              & 7                                                                         & $\Ltwo$           & 4759                                                                  & 6.22              & 6.3           & 8             & 427.07                                                                & 407.66                                                                            & \textbf{1.05} \\ \hline
15                                                                              & 7                                                                         & $\Ltwo\&\Lone$        & 5552                                                                  & 12.09             & 22.49         & 29.26         & 388.3                                                                 & 381.44                                                                            & 1.02          \\ \hline
15                                                                              & 7                                                                         & $\Lone$           & 5132                                                                  & 6.02              & 5.86          & 7.57          & 427.44                                                                & 405.72                                                                            & \textbf{1.05} \\ \hline
15                                                                              & 10                                                                        & $\Ltwo$           & 6409                                                                  & 7.35              & 6.11          & 7.58          & 432.25                                                                & 412.92                                                                            & \textbf{1.05} \\ \hline
15                                                                              & 10                                                                        & $\Ltwo\&\Lone$        & 4354                                                                  & 10.5              & 15.77         & 20.16         & 393.55                                                                & 380.94                                                                            & 1.03          \\ \hline
15                                                                              & 10                                                                        & $\Lone$           & 5002                                                                  & 7.11              & 6.49          & 8.64          & 423.17                                                                & 402.73                                                                            & \textbf{1.05} \\ \hline
\end{tabular}
\vspace{0.5cm}
\end{table*}

\begin{table*}[!t]
\centering
\caption{Effect of the Learning rate $\alpha$ and discount factor $\gamma$.}
\label{tab-alpha-gamma}
%
\begin{tabular}{|c|c|c|c|c|c|c|c|c|}
\hline
$\mathbb{\alpha}$ & $\mathbb{\gamma}$ & \textbf{\begin{tabular}[c]{@{}c@{}}~Training~\\ ~~~episodes~~~~\end{tabular}} & ~~\textbf{MAPE (\%)}~~ & ~~\textbf{MAE}~~   & ~~\textbf{RMSE}~~  & \textbf{\begin{tabular}[c]{@{}c@{}}~~Total food~~\\ ~~(grams)~~\end{tabular}} & \textbf{\begin{tabular}[c]{@{}c@{}}~~Final fish~~\\ ~~~weight (grams)~~~~\end{tabular}} & ~~~\textbf{FCR}~~~  \\ \hline
~~0.1~~          &~~0.1~~        & 8001          & 11.01         & 21.75         & 29.10         & 373.34          & 362.16          & 1.03          \\ \hline
0.1          & 0.5        & 7312          & 6.08          & 7.85          & 10.06         & 431.82          & 418.41          & 1.03          \\ \hline
0.1          & 0.7        & 6748          & 5.25          & 5.37          & 6.87          & 447.60          & 430.65          & 1.04          \\ \hline
\textbf{0.1} & \textbf{1} & \textbf{9171} & \textbf{4.31} & \textbf{4.20} & \textbf{5.32} & \textbf{449.58} & \textbf{428.75} & \textbf{1.05} \\ \hline
0.5          & 0.1        & 7014          & 12.84         & 20.29         & 27.48         & 390.46          & 374.38          & 1.04          \\ \hline
0.5          & 0.5        & 6572          & 5.27          & 5.55          & 7.21          & 434.87          & 417.58          & 1.04          \\ \hline
0.5          & 0.7        & 6455          & 5.00          & 4.51          & 5.71          & 448.74          & 428.75          & 1.05          \\ \hline
0.5          & 1          & 5968          & 5.65          & 4.68          & 5.76          & 441.60          & 419.94          & 1.05          \\ \hline
\end{tabular}
\vspace{0.5cm}
\end{table*}

\begin{table*}[!t]
\centering
\caption{Effect of the exploration phase duration $t_{\epsilon}$.}
\label{tab-eps}
\begin{tabular}{|c|c|c|c|c|c|c|c|}
\hline
\textbf{$t_{\epsilon}$} & \textbf{\begin{tabular}[c]{@{}c@{}}Training \\ ~~~~episodes~~~~\end{tabular}} & ~~\textbf{MAPE (\%)}~~ & ~~\textbf{MAE}~~  & ~~\textbf{RMSE}~~ & \textbf{\begin{tabular}[c]{@{}c@{}}~~Total food~~\\ (grams)\end{tabular}} & \textbf{\begin{tabular}[c]{@{}c@{}}~~~Final fish~~~ \\ ~~~weight (grams)~~~\end{tabular}} & \textbf{~~~FCR~~~}   \\ \hline
~~3000~~                    & 26459                                                                 & 5.56              & 5.38          & 6.73          & 434.61                                                                & 415.77                                                                       & 1.045          \\ \hline
\textbf{6000}           & \textbf{15294}                                                                 & \textbf{5.16 }             & \textbf{5.25}          & \textbf{6.73}          & \textbf{431.43}                                                       & \textbf{413.01}                                                              & \textbf{1.045} \\ \hline
9000           & 11104                                                        & 5.16     & 5.74 & 7.53 & 436.12                                                                & 417.80                                                                       & 1.044          \\ \hline
\end{tabular}
\vspace{0.5cm}
\end{table*}

\subsubsection{Discretization Scheme  and Reward}\label{senst1}
The discretization scheme plays an essential role in defining a finite state set based on converting the continuous dynamics to finite/discrete dynamics. In the analysis, two different discretization resolutions of the fish weight and age are compared. In addition, three different state-action rewards are adopted as follows
\begin{equation*}
\Ltwo:\quad R(s_t,a_t)= - \left(\dfrac{ w(s_t)- w^d(t)}{w^d(t)}\right)^2 - \lambda \Big(f_{a_{t}}\Big)^2,
\end{equation*}
\begin{equation*}
\Ltwo\&\Lone: \quad R(s_t,a_t)= - \left(\dfrac{ w(s_t)- w^d(t)}{w^d(t)}\right)^2 - \lambda \abs{f_{a_{t}}},
\end{equation*}
\begin{equation*}
\Lone:\quad R(s_t,a_t)= - \Big|\dfrac{ w(s_t)- w^d(t)}{w^d(t)} \Big| - \lambda \abs{f_{a_{t}}},
\end{equation*}
where $w(s_t)$ is the fish weight at the state $s_t$, $w^d(t)$ is the desired fish weight at the time $t$ and $f_{a_{t}}$ is the feeding rate. $\lambda$ is the regularization term used to assess overfeeding and feeding consumption. The advantage of using $\Lone$-norm and the combination of $\Ltwo$ and $\Lone$ norm is to smoothly reduce the RL control strategy due to the increasing behavior of the reward function.  Table \ref{descritsation_resolution} shows by decreasing the resolution improves the tracking performance and increasing the training/learning duration. Besides, Table \ref{descritsation_resolution} shows that $\Ltwo$  reward seems to give the highest FCR and training time duration.

\subsubsection{Learning Rate $\alpha$ and Discount Factor $\gamma$}\label{senst2}
Table \ref{tab-alpha-gamma} demonstrates that decreasing the learning rate $\alpha$ and increasing the discount factor $\gamma$ improve the tracking performance and reduce the needed time for learning by the number of learning episodes. 


\subsubsection{Exploration and Exploitation $\epsilon$}\label{senst3}
The exploration phase is a crucial step in the learning process. It allows the agent to observe the different possible scenarios and helps in improving the performance. However, exploring too much might mislead the agent and can lead to a slow convergence rate. Consequently, an optimal choice of the exploration duration is needed by the optimal choice of the value of the parameter $\epsilon$ as defined in equation \eqref{greddy_annealing}. Table \ref{tab-eps} shows that increasing the exploration phase duration improves the learning by providing good performance in shorter training duration.

\section{Conclusion}\label{sec-dis-con}
This paper proposes a model-free reinforcement learning-based control policy to achieve these goals and meet the precision aquaculture target.  The proposed Q-Learning based control strategy is applied to the two simulated types of aquaculture systems, whether for floating cages or tanks on land.  The obtained results show that  Q-learning feeding and temperature control policy can achieve the desired growth rate while reducing the feed quantity. For instance, it can correctly track an experimental fish growth,  collected from a research project in Thailand \cite{Dampin2012}, with an increase of  $11\%$  in the feeding quantity compared to the ideal cases where the water temperature is maintained to its optimal value.  The Q-learning policy demonstrates a remarkable ability to learn the dynamic from collected and build a suitable control policy. 
Overall, these simulated trained Q-learning policies can be used to start a new learning phase without exploration in real experiments aquaculture environment. Additionally, the Q-learning controller can be improved further by considering a larger learning problem by adding more states and actions. 

In the future, the proposed Q-learning policies will be combined with a safe-learning framework to train it in a real environment further. This will be the first step toward deploying this learning algorithm in a real aquaculture environment.
The Q-learning implementation is publicly available online and downloadable from: https://github.com/EMANG-KAUST/Q-Learning-in-aquaculture.git. 

\section*{Acknowledgment}
The authors would like to thank Professor Jeff Shamma from King Abdullah University of Science and Technology (KAUST) for helpful discussions and guidance on reinforcement learning.

\balance
\bibliography{ref}

\end{document}